# Deformation mechanisms in PBT at elevated temperatures

Laurent FARGE[1], François TOURNILHAC[2], Javier PEREZ[4], Sandrine HOPPE[3], Isabelle BIHANNIC[5], Jérémy BIANCHIN[1], Julien BOISSE[1], Stéphane ANDRÉ[1]

1 Université de Lorraine, CNRS, LEMTA, F-54000 Nancy, France
2 Molecular, Macromolecular Chemistry, and Materials, ESPCI Paris, PSL Research University, 10 Rue Vauquelin, F-75005 Paris, France
3 Université de Lorraine, CNRS, LRGP, F-54000 Nancy, France
4 Synchrotron SOLEIL, L'Orme des Merisiers, Saint-Aubin, BP 48, 91192 Gif-sur-Yvette Cedex, France
5 Université de Lorraine, CNRS, LIEC, F-54000 Nancy, France

## Abstract

The plastic deformation of Polybutylene terephthalate (PBT) is investigated at three temperatures (120°C, 150°C, 180°C) representative of the span between glass transition, $T_g \approx 55$°C and melting $T_m \approx 225$°C temperatures. The mechanisms of tensile deformation are analyzed as a function of the true strain using a combination of temperature–controlled tensile experiments coupled with 2D digital image correlation and synchrotron sourced wide angle (WAXS) and small angle (SAXS) X-ray scattering measurements.

Plots of the force signal in nominal stress vs true strain, true stress vs true strain and Haward-Thackray representation reveal the same three different regimes I, II, III and their respective strain ranges are found virtually independent of temperature. Between $\varepsilon = 0$ and the yield strain observed at about $\varepsilon_y \approx$ 0.3 (strain range I) the layer morphology and average crystalline thickness are preserved but the interlamellar distance increases in the drawing direction; the lamellar correlation signal becomes anisotropic but the strain is localized in amorphous regions where the chains remain describable as gaussian coils. Between $\varepsilon \approx 0.3$ and $\varepsilon \approx 1.0$ (strain range II) the gaussian approximation still holds but the lamellae begin to fragment, as revealed by a shortening of the correlation distance, except at 180°C where melting/recrystallization processes take place. According to WAXS data, strain range II corresponds to the transition from spherulitic to fibrillar morphologies. Beyond $\varepsilon \approx 1.0$ (strain range III) the chains are taut and the fibrillar morphology is established, any additional stress develops in both the amorphous and crystalline regions causing the transition between polymorphs, respectively from the isotropic amorphous phase to a smectic phase and from the α crystalline phase to the β phase.

1 Introduction

Polybutylene terephthalate (PBT) is an engineering polymer, remarkable by its relatively high melting point ($T_m \approx 225$°C), low melt viscosity, fast crystallization and low water uptake, which makes it particularly suitable for implementation by injection molding and use in electric insulation, cooking tools, electric appliances, composites etc. One of the key properties in all these applications is the mechanical resistance at high temperature. Efforts have been made in recent years to improve the mechanical resistance of PBT both above and below the melting point by chemical modification along the concept of vitrimers [Demongeot (2016), Farge (2020)]. Another important topic to progress in this direction is the understanding of plastic deformation in different ranges of temperature, with respect to $T_g$ and $T_m$.

Semi-crystalline polymers subjected to tensile drawing above their glass transition temperature ($T_g$) can be easily deformed until reaching the neck stabilization regime [Ward (2013), Galeski (2003)]. The strain in the neck becomes then approximately constant leading to large values, roughly in the range 200 %–1000 % (or true strains 0.7–2.3) depending on the semi-crystalline polymer under study. On the other hand, when the test is performed at temperatures lower than that of the glass transition, the polymer exhibits a "brittle behavior" and rupture occurs at much lower strains. In the former case — test performed above the glass transition and until large final strains — drastic transformations of the polymer microstructure occur during the test corresponding to the change of the initial lamellar morphology into fibrillar morphology. Since the end of the 1990s, the details of these microstructure transformations have been thoroughly studied using *in situ* synchrotron sourced WAXS (Wide Angle X-ray Scattering) and SAXS (Small Angle X-ray Scattering) experiments. These experiments were initially carried out at room temperature. This means that it was not possible to study the effect of temperature on the deformation process, and that it was conceivable to study the microstructure transformations occurring at large strains only in the case of semi-crystalline polymers whose $T_g$ is lower than the room temperature. As a result, there exists a large number of papers concerning polyethylene ($T_g \approx$ -110°C) [Hughes (1997), Butler (1995, 1997, 1998), Yakovlev (2019), Lin (2019)], and a few less related to polypropylene ($T_g \approx$ -10°C) [Davies (2003), Qiu (2007)]. Since the beginning of the 2010s, several *in situ* SAXS/WAXS studies were published reporting tests carried out above room temperature [Mao (2013), Paolucci (2019), Xiong (2013), Xiong (2015), Chen (2017), Cai (2012), Lin (2019)]. This is experimentally more difficult since this requires using a heating device coupled to a tensile machine in order to perform the test *in situ* on an appropriate synchrotron beamline.

In the case of PBT ($T_g \approx$ 55°C), we found only one study by Zhang et al. where the large-strain tensile behavior of this polymer was studied by means of *in situ* SAXS/WAXS measurements [Zhang (2018)]. During tensile testing, the authors highlighted three different stages during the deformation process, each associated with specific deformation mechanisms. The local true strain was not measured in the neck center, that is to say where the SAXS/WAXS measurement is carried out. Only engineering strain values — a priori deduced from the tensile machine crosshead relative displacement and therefore not intrinsic to the material — were provided to indicate the thresholds separating these three stages. The highest temperature for which SAXS/WAXS measurements were reported was 120°C, which is significantly smaller than the melting temperature ($T_m \approx$ 25°C) and leaves a wide range of temperatures not investigated.

With exception of this reference, there is no SAXS/WAXS data about the deformation process and microstructure evolution of PBT during tensile drawing to large strains at temperatures above $T_m$. However, the tensile behavior of oriented PBT at room temperature has been widely studied by WAXS [(Desborough (1977), Carr (1997), Yokouchi (1975), Bereton (1977), Jakeways (1975), Tashiro (1980)]. It was found that a reversible phase transition between the α and β triclinic forms occurs at strains between 4% and 12%. This transition proceeds from a change in the glycol residue conformation from the gauche-trans-gauche sequence (α form) to a nearly all‑trans sequence (β form). As a result, the *c* parameter (fiber axis) of the triclinic unit cell significantly changes during the α → β transition (from *c* = 11.67 Å in the α-form to *c* = 12.90 Å in the β form) while the five other parameters remain nearly unchanged. The change of the *c* parameter is discontinuous, no intermediate state has been reported [Jakeways (1975)]. In the β phase — all-trans conformation sequence for the glycol residue — the polymer chains become very close to full extension.

Smectic structures have also been observed in amorphous PBT and in other amorphous polyesters — obtained from molten state by quick quenching — subjected to stretching below $T_g$; for example PBT [Song(2000),Konishi (2010)], PET (polyethylene terephthalate) [Kawakami(2005), Sago(2014), PEN (polyethylene naphtalate) [Jakeways (1996)].

In this study, we have carried out synchrotron *in situ* X-ray scattering experiments to analyze the deformation mechanisms of a non-oriented PBT subjected to tensile drawing to large strains. A special heating device was designed to carry out measurements at elevated temperatures (180℃). Complementary optical experiments were performed to analyze the results in terms of local true strains measured in the specimen center, precisely where the X-ray beam passes through the specimen. The approach is therefore significantly different from that of Zhang et al. [Zhang (2018)], the sole study on the same subject. True strain analysis on a broad temperature range enabled us to measure the threshold strains between different plastic regimes and to provide many new results and interpretations. In particular, orientational effects of the α–β transition and the appearance of a smectic structure were observed at large strains, in conditions not previously reported.

# 2 Experimental section

## 2.1 PBT production and specimen shape

The PBT studied in this work was provided by DuPont in pellet form (Material reference: CRASTIN FGS600F40). The specimens were produced in a DSM Xplore μcompounder at 270°C. The rotational speed of the two μcompounder screws was 100 rpm. At the μcompounder outlet, the melt was transferred to a shooting pot and injected into a mold in shape of a tensile specimen (see Figure S1 of Supplementary Data, appendix B). A 4 mm length flat part was machined on the specimen lateral faces to promote a uniaxial stress state in the center of the specimen (see Figure S2 for specimen dimensions). Using this geometry, the necking develops in the specimen central region, exactly where the X-ray scattering measurement is carried out. During a previous study [Farge (2021)], DSC (Differential Scanning Calorimetry) measurements were performed on the same PBT and the measured glass temperature, melt temperature and crystallinity were found to be 50°C, 225°C and 34% respectively.

## 2.2 *In situ* X-ray scattering measurements during tensile testing

### 2.2.1 Experimental set-up

The experiments were carried out using a Kammrath & Weiss mini-machine specifically designed for *in situ* X-ray scattering studies with synchrotron radiation. The two mini-machine crossheads move apart from each other at constant speed $\dot{u} = 20$ μms⁻¹ so that the displacement of the central specimen point — where the X-ray beam passes through the specimen — is zero. To perform high-temperature tensile tests, it was not possible to put the entire tensile machine in a heat chamber because the machine electrical and electronic components (controlling device, sensors) are not intended to work at temperatures required for our experiments (up to 180°C). We were therefore forced to design a specific heating device for this study. It consists of a heating resistance included in a small metal part that surrounds the specimen central region (see Figure 1). As a result, only the specimen central part is subjected to high temperatures, which ensures the proper functioning of the tensile machine and of its electronics. Two apertures were cut in the heating device in front and rear faces to let the X-ray beam pass through the specimen. The tensile mini-machine was oriented horizontally to avoid gravity natural convection, which might make necking to initiate above the specimen center.

The tensile tests were interrupted when the displacement between the two crossheads was 15 mm. The deformation state in the specimen center was characterized by the value of the true longitudinal strain: $\varepsilon = \ln \lambda$ . $\lambda$ is the draw ratio: $\lambda = \ell / \ell_0$, where $\ell_0$ is a small length element taken in the specimen center along the tensile axis in the initial state. $\ell$ is the current size of this length element during tensile testing. The nominal stress is given by $\sigma_N = F / S_0$ , where $S_0$ is the 4×4 mm$^2$ area of the central cross-section. Assuming an isochoric deformation process, the true stress that accounts for the decreasing cross-sectional area is given by: $\sigma_V = \sigma_N e^{\varepsilon}$ .

By using literature data [Huo(1992), Desborough(1977)] for the unit cell parameters of the two triclinic phases, it is easy to find that the density of the crystalline parts is decreased by about 5% during the α–β transition (see paragraph SD1 for more details in the Supplementary Data file, appendix B). For a 34% crystallinity, this corresponds to a 1.7% decrease of the polymer overall density. We checked that the error on $\sigma_V$ due to the assumption of an isochoric deformation process is then smaller than 2%.

Prior to synchrotron experiments and at all targeted temperatures, 2D Digital Image Correlation (2D DIC) experiments were carried out through the heating device aperture in order to measure the strain evolution $\varepsilon(t)$ during the tensile tests (Figure 2). The $\varepsilon(t)$ curves are found to depend little on temperature even though the final quasi-constant strain is slightly higher at 180°C. These results were confirmed in an independent way through the analysis of the transmission signal that was measured during the X-Ray scattering experiments. During the test, the transmission increases due to thickness reduction. Using the Beer-Lambert law and the assumption of transversally isotropic and isochoric deformation process, a second evaluation of $\varepsilon(t)$ was achieved, from the evolution of the transmission signal. A satisfactory agreement was found between both (see Figure S3), which supports the approximation of isochore deformation is valid. In this study only the $\varepsilon(t)$ values obtained by DIC measurements will be used.

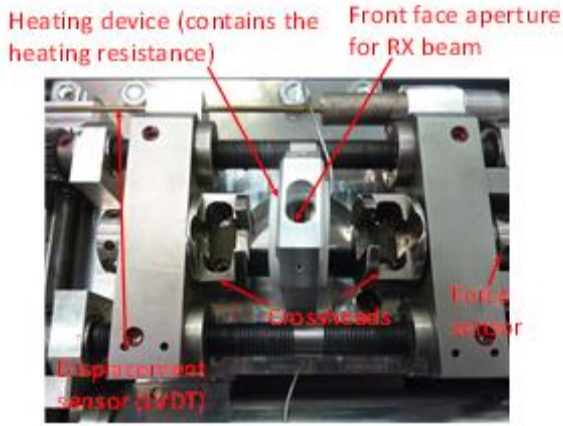

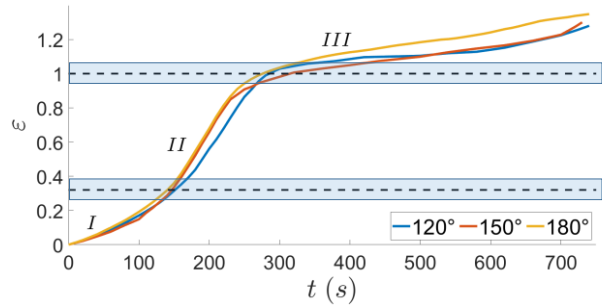

*Figure 1 Tensile mini-machine equipped with a heating device specifically dedicated to synchrotron studies. (A high-resolution image will be provided if needed).*

*Figure 2 Evolution of the true longitudinal strain $\varepsilon(t)$ measured in the specimen central cross-section*

## 2.2.2 X-ray scattering measurement

The X-ray scattering measurements were performed on the SWING beamline at the Soleil synchrotron. The radiation wavelength was 0.77489 Å (16 keV). Measurements were made at a sample-detector distances of 1244 mm for three test temperatures: 120°C, 150°C and 180°C. 75 patterns were *in situ* recorded every 10 seconds during a single tensile test with a 1 s acquisition time. An example of pattern measured at 120°C and corresponding to the initial undeformed state is shown in Figure 3a. The lamellar ring is visible in the right-down part of the image corresponding to the "SAXS part" of the pattern. The detector position was chosen so that a full quadrant — azimuthal angles ($\varphi$) ranging from 0° to 90° — is covered for the (001) reflection. A part of the (011) and (010) reflection can also be seen in upper-left part of the pattern.

In this study, we will mainly analyze the (001) reflection, and the evolution of the correlation distance corresponding to the lamella periodic organization. In both cases, the Bragg angle $\theta$ is very small: $\theta < 0.25°$ for the lamellar periodic morphology, and $\theta \approx 2°$ for the (001) reflection. The Polony relation:

$\cos\psi_L = \cos\theta\cos\varphi$ (see Figure 3b) — linking the Bragg angle $\theta$, the angle $\psi_L$ between the tensile direction and the normal to the lattices ($\vec{n}_{\psi_L}$) and the azimuthal angle $\varphi$ defined on the scattering pattern (Figure 3a) — leads to consider that $\psi_L \approx \varphi$.

In the SAXS part of the pattern, the evolution of the correlation distance associated with lamella scattering was measured along the drawing direction both with ($d^{Ltz}$) and without ($d$) the Lorentz correction. To obtain $d^{Ltz}$, we first determined the position ($q_{max}$) corresponding to the local maximum of $q^2 I(q)$ that is specifically associated with X-ray lamella scattering in the $\varphi = \begin{bmatrix} 0° & 10° \end{bmatrix}$ angular sector, i.e. along the drawing axis. Due to image sampling by pixels, only discrete values are possible for $q_{max}$. The final value for the maximum position $q'_{max}$ is found by interpolating the data around $q_{max}$ with a second order polynomic function. $d^{Ltz}$ is finally given by: $d^{Ltz} = 2\pi / q'_{max}$. To obtain the correlation distance without the Lorentz correction ($d$), the same procedure was carried out but for determining the maximum of the $I(q)$ profile (instead if $q^2 I(q)$).

Based on a two-phase model of the lamellar morphology and using the procedure described by Strobl et al. [Strobl (1980)] (illustrated in Figure S4 and detailed in SD2), the proportion of crystalline phase was extracted from the isotropic SAXS patterns measured at the initial state by analyzing the characteristics of the correlation function $K(r)$ that is defined as follows:

$$K(r) = \frac{\int_0^\infty I(q)q^2 \cos(qr)dq}{\int_0^\infty I(q)q^2 dq} \qquad \text{Equation 1}$$

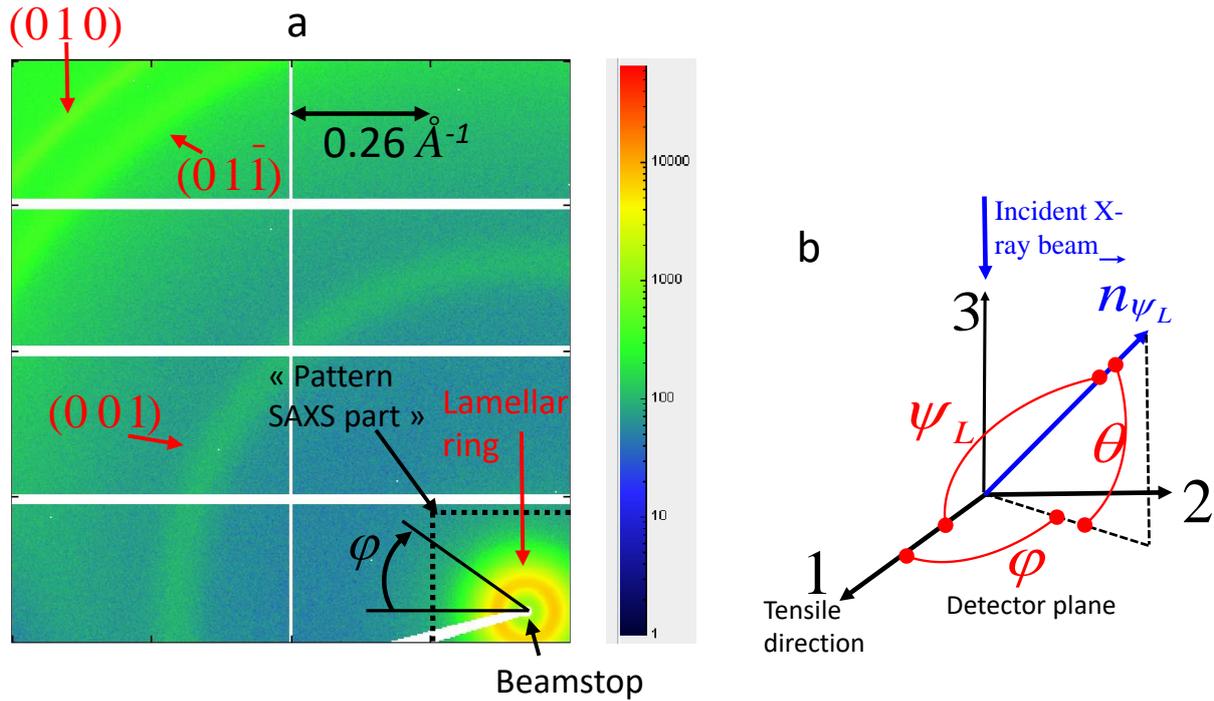

*Figure 3 X-ray scattering experiment. a) Example of X-ray pattern measured at 120°C and at the initial state ( $\varepsilon = 0$ ) and b) Scheme of the experiment geometry.*

In order to identify the different possible crystalline phases for PBT ($\alpha$ or $\beta$ forms), complementary measurements were performed at 150°C using a sample-detector distance of 726 mm (see Figure S5a for the pattern recorded before deformation). In addition to the (001), (01$\bar{1}$) and (010) reflections, parts of the (1$\bar{1}$1), (1$\bar{1}$0), (100) and (11$\bar{1}$) reflections were also identified on the patterns. Unfortunately, it was not possible to obtain the reflection intensities on a complete quadrant. At the heating device outlet, the aperture dimension along the tensile direction is too small to allow the photons to go out if the scattering angle is too large. Incidentally, this blind sector is also the region where uniaxially oriented crystals are not expected to produce diffraction signals, because nodes located along the drawing direction, especially at large $q$ do not intercept the Ewald sphere anymore when orientation is strong.

# 3 Results and discussion

## 3.1 Tensile testing

In Figure 4, we show the evolution of the nominal stress $\left( \sigma_N \right)$ versus the true strain $\left( \varepsilon \right)$ for tensile tests performed at 120°C, 150°C and 180°C. These three curves exhibit three different stages which start and end at well defined values of the true strain, approximately independent of temperature. These three stages can be described as follows. From $\varepsilon = 0$ to $\varepsilon \approx 0.3$ (strain range I), the stress strongly increases until the yield point. The yield strain $\left( \varepsilon_Y \right)$ is approximately the same for the three test temperatures ($\varepsilon_Y \approx 0.29$ at 120°C, $\varepsilon_Y \approx 0.32$ at 150°C, $\varepsilon_Y \approx 0.35$ at 180°C). From $\varepsilon \approx 0.3$ to $\varepsilon \approx 1.0$ (strain range II), the nominal stress only shows small variations. Note that in the figures of the article, the line indicating the end of strain range I is placed at $\varepsilon \approx 0.32$. Strain range III begins approximately at $\varepsilon \approx 1.0$ for the three test temperatures and corresponds to an abrupt increase of $\sigma_N$. It is also interesting to analyze these tensile tests using the true stress $\sigma_V$ that correctly accounts for the decreasing cross-sectional area. Different representations are possible. In Figure 5, the true stress $\sigma_V$ is plotted as a function of $HT = \lambda^2 - 1/\lambda = e^{2\varepsilon} - e^{-\varepsilon}$ which is the representation associated with the Haward-Thackray model. [Haward (1968), Haward (1993)]. In this representation, the slope of the linear portion of the curve (above the yield point) is related to the elastic shear modulus $G_p$ through the relation: $\sigma_V = Y + G_p\,(e^{2\varepsilon} - e^{-\varepsilon})$. Though the test was not carried out at constant strain speed, we can still use this model for qualitatively interpreting our results without expecting the determination of an accurate value for the network shear modulus. Despite a noticeable distortion of the plot, the transitions between regimes I, II and III are detected for the same values of true stress (Figure 5). Unlike $\sigma_N$ that is roughly constant during strain range II, $\sigma_V$ increases with $e^{2\varepsilon} - e^{-\varepsilon}$ in an approximately linear way giving rise to shear moduli in the range of a few MPa. Such a behavior has been commonly observed for semi-crystalline polymers [Haward (1993)]. According to the Haward-Thackray approach, this means that the Gaussian chain approximation is valid in describing the macromolecular network stretching accompanying the plastic instability flow. During strain range III, the previous linear regime is finished and the true stress strongly increases, which shows that the Gaussian chain approximation is no longer valid. The network is fully extended, which occurs when the polymer chains are taut. Unwinding part of the chains from the crystals is then necessary to allow further deformation. Another representation is plotted in Figure S6 (true stress as a function of the true strain) where the same three stages of the deformation process are identified. Interestingly, these three stages, separated by the same strain thresholds ($\varepsilon \approx 0.30$ and $\varepsilon \approx 1.0$) also appear on the $\varepsilon(t)$ curves in Figure 2 or S3. During strain range II, after the yield point, the strain rate strongly increases. During strain range III, the strain increase becomes very small in the neck center ($\approx$ specimen center) because strain hardening is very strong in this zone (see Figure S6). The plastic instability migrates in other specimen regions, namely the two symmetrical neck shoulders [Ye (2015), Farge (2018)].

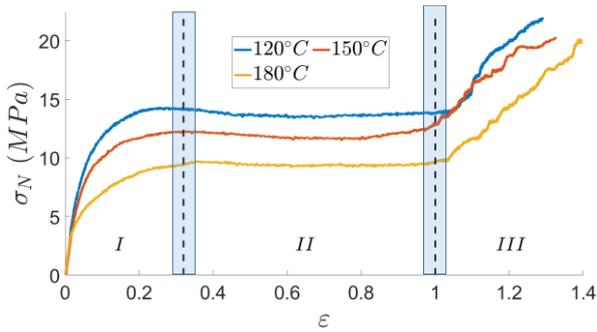

*Figure 4 Evolution of the nominal stress: $\sigma_N(\varepsilon)$ curves*

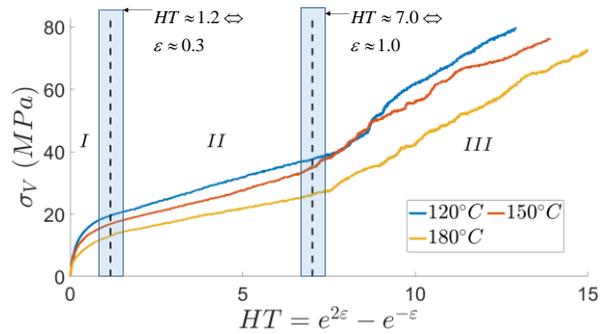

*Figure 5 True stress $\sigma_V$ versus $e^{2\varepsilon} - e^{-\varepsilon}$ (Haward-Thackray representation)*

## 3.2 SAXS analysis

### 3.2.1 Orientation of the SAXS patterns

Examples of SAXS patterns (extracted from the down-right part of the full detector plane as shown in Figure 3a) are given in Figure 6 at various strain levels and for the three test temperatures.

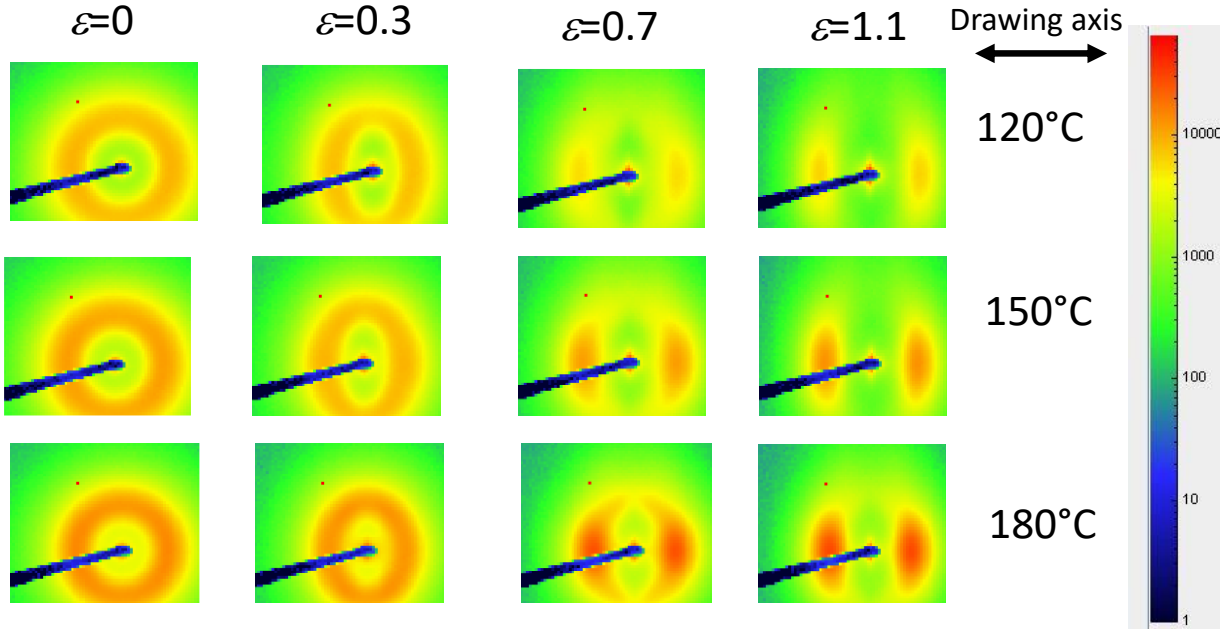

*Figure 6 Examples of SAXS patterns*

As expected, we can observe the circular ring associated with the isotropic lamellar morphology at the initial state ($\varepsilon = 0$). The proportion of crystalline phase extracted from the correlation function are 28.5% at 120°C, 30.1% at 150°C and 30.7% at 180°C. These values are between 10% and 19% smaller than the crystallinity measured by DSC. At the yield point ($\varepsilon \approx 0.3$), the ring is deformed because the correlation distance increases for the lamellae perpendicularly oriented with respect to the drawing axis while it decreases for the lamellae oriented along this direction. However, the intensity along the ring remains approximately constant. At larger strains ($\varepsilon = 0.7$) the lamellar ring is hardly discernible and the main feature of the SAXS patterns becomes the two intensity maxima along the horizontal axis (drawing axis) that are characteristics of the fibrillar morphology. Finally, at $\varepsilon = 1.1$, the fibrillar morphology is definitely established.

In order to assess quantitatively the progression of the lamellar/fibrillar transition as a function of $\varepsilon$, an orientation parameter $A$ was calculated as follows:

$$A = \frac{I_{SAXS}^{H} - I_{SAXS}^{V}}{I_{SAXS}^{H} + I_{SAXS}^{V}} \qquad\qquad \text{Equation 2}$$

$I_{SAXS}^{H}$ and $I_{SAXS}^{V}$ are the average intensities in angular sectors corresponding respectively to $\varphi = \begin{bmatrix} 0° & 10° \end{bmatrix}$ and $\varphi = \begin{bmatrix} 80° & 90° \end{bmatrix}$, and limited to a q range that includes almost all the intensity specifically scattered by lamellae ($q = \begin{bmatrix} 0.035 & 0.2 \end{bmatrix} Å^{-1}$). $A = 0$ in the case of the isotropically disoriented lamellar morphology. When the fibrillar morphology is being formed, the intensity of the lamellar peak concentrates along the horizontal axis, which leads to an increase of $A$. The three regimes (strain ranges *I*, *II* and *III*) identified on the stress-strain curves of Figure 4, 5 and S6 measured on the macro-scale are also clearly visible on the $A(\varepsilon)$ curves (Figure 7). In particular, a strong increase of the orientation parameter occurs during strain range *II*. This suggests that the lamellar/fibrillar transition occurs mainly during this strain interval. At the beginning of strain range *III* (for $\varepsilon > 1.0$), the strong increase of the orientation parameter is finished and $A$ becomes nearly constant; the higher the temperature, the higher the final $A$ value. At small strains ($\varepsilon < 0.3$), after a slight initial rise, $A$ moderately decreases. This may be caused by the formation of a "kink morphology", which corresponds to local disorientation of the lamellae with normals initially oriented along the drawing axis. The occurring of kink (or chevron) morphology in the spherullite equatorial parts has often been reported [Krumova (2006), aleski (1988), Pawlak (2014)]. This can be considered as a form of buckling due to the existence of a compressive stress along the lamellae of these regions.

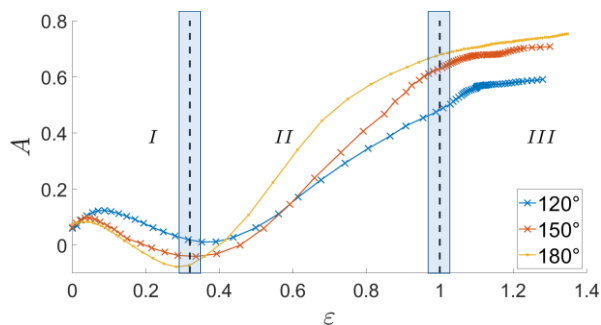

*Figure 7 Orientation parameters extracted from the SAXS patterns.*

### 3.2.2 Analysis of the correlation distance evolution

The evolution of the correlation distance ($d^{Lrz}$ with the Lorentz correction) along the drawing direction is plotted in Figure 8. This quantity associated with nano-scale is again analyzed in the framework of the three previously identified regimes at macro-scale.

*Evolution of the correlation distance during Stage I*

As it has often been observed for semi-crystalline polymers [Jiang (2009), Zhang (2018), Stribeck (2003), Chang (2018), Lifu (2019), Habumugisha (2021), Liao (2017)], $d^{Lrz}$ begins to increase. The lamellae's integrity is first overall preserved and the lamellae normally oriented with respect to the tensile direction move away from each over, which causes the observed $d^{Lrz}$ increase. Precisely at the transition between strain ranges *I* and *II* ($\varepsilon \approx 0.3$), $d^{Lrz}$ reaches a maximum.

*Stage I/Stage II Transition and evolution of the correlation distance during stage II*

Next, $d^{Lrz}$ significantly decreases from the beginning of strain range *II* at 120°C and 150°C. The lamellae's fragmentation is in progress, which leads to an increase of the number of crystalline regions per volume unit, and logically to the decrease of the correlation distance ($d^{Lrz}$). On the other hand, the evolution of $d^{Lrz}$ during strain range *II* differs drastically at 180°C: instead of decreasing, the correlation distance remains nearly constant. Jiang et al. have obtained very comparable curves for the dependence of the $d^{Lrz}(\varepsilon)$ curves on temperature in the case of a high-density polyethylene. They showed that during solidification from the melt, certain crystal blocks are likely to crystallize at temperatures lower than that of the tensile test carried out afterwards. During this tensile test, these crystal blocks become unstable because they are subjected to stress and melt before rapidly recrystallizing at the test temperature with higher thicknesses. This leads to an increase in $d^{Lrz}$, which may compensate the decrease trend due to fragmentation. This is what occurs for the test performed on PBT at 180°C (Figure 8). On the other hand, the number of crystal block formed below 150°C during solidification is too small for the decreasing trend of $d^{Lrz}$ caused by fragmentation to be compensated in the case of the test carried out at this temperature.

As it was already mentioned, it is possible to extract the lamella thicknesses ($d_c$) at the initial undeformed state from the correlation function $K(r)$ (Figure S4). We found that $d_c$ is 28.2 $\mathring{A}$ at 120°C, 31.1 $\mathring{A}$ and 35.3 $\mathring{A}$ at 180°C. The increase of $d_c$ with temperature — more pronounced in the 150-180°C range — suggests that the aforementioned melting/recrystallization process is likely to occur even if no stress is applied.

As proposed by Zhang (2018) and Mao (2018), we have calculated from the correlation function $K(r)$ (Equation 1) and plotted in Figure 9 the evolution of both the crystal $d_c(\varepsilon)$ and amorphous $d_a(\varepsilon)$ thicknesses as function of the applied strain. The reliability of such a calculus can be discussed because the Lorentz correction assumes an isotropic material, which is not true during the tensile process. But as shown in Figure 7, the bias due to anisotropy development is nearly the same at all three temperatures and consequently we believe that the $d_c(\varepsilon)$ curves difference in behavior is relevant for analyzing the temperature effect on the deformation process. Notably, referring to figure 9 we can observe that $d_c(\varepsilon)$ increases during strain range *II* at the three test temperatures and especially as the temperature is elevated. This may confirm the aforementioned discussion about the results of Figure 8 during strain range *II*, namely the melting of crystal blocks due to stress and rapid recrystallization with higher thicknesses at the test temperature.

*Stage II/Stage III transition and evolution of the correlation distance during strain range III*

The transition at $\varepsilon \approx 1.0$ seems to correspond to a slight increase of the $d^{Lz}(\varepsilon)$ curves (Figure 8). Nevertheless, this is not easy to assure because $d^{Lz}(\varepsilon)$ seems increasing from about $\varepsilon \approx 0.8$. At these strains the polymers are strongly oriented (Figure 7) and the use of the Lorentz correction is highly questionable. Therefore, we plot in Figure 10 the correlation distance *calculated now* without Lorentz correction ($d$) and only in the "large strains range" ($\varepsilon > 0.7$). The strain threshold previously highlighted at $\varepsilon \approx 1$ is well confirmed here. The $d(\varepsilon)$ evolution on the whole strain range is shown in Figure S7.

*Summary*

The analysis of the correlation distance evolution helps to understand the changes of microstructure associated with the three stages of the deformation process that were identified from the macro-level in the stress-strain curves (Figures 4, 5 and S6). During strain range $I$ ($\varepsilon \approx \begin{bmatrix} 0 & 0.30 \end{bmatrix}$), the observed increase in correlation distance is due to the rise of the spacing between adjacent lamellae with normal along the drawing axis, the lamellar integrity is roughly preserved and the strain is localized in amorphous layers. During stage $II$ ($\varepsilon \approx \begin{bmatrix} 0.30 & 1 \end{bmatrix}$), the lamella fragmentation occurs and the number of crystal zones per volume unit increases, which results in the decrease of the correlation distance. However, at 180°C, the stress induced melting of unstable crystals, which quickly recrystallize with higher thicknesses, can compensate this effect. During stage III, the lamella fragmentation is finished and the fibril morphology is established. As in stage $I$, the strain is concentrated in amorphous regions and the correlation distance resumes increasing.

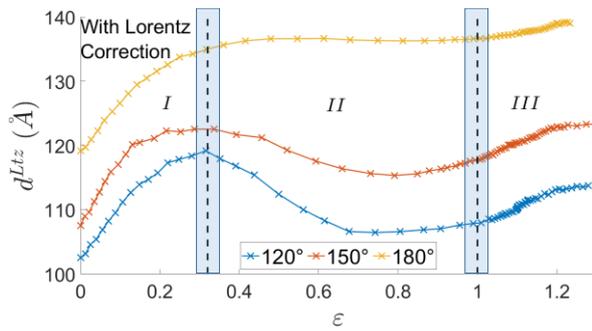

*Figure 8 Evolution of the correlation distance along the drawing axis calculated using the Lorentz correction.*

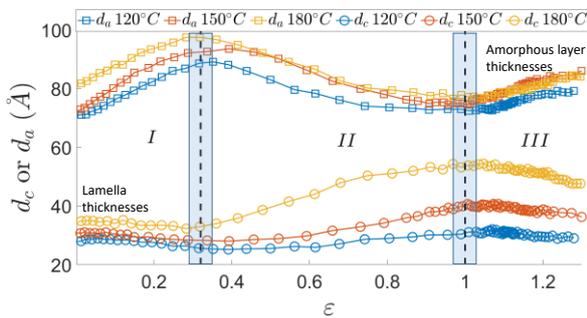

*Figure 9 lamella and amorphous layer thicknesses along the drawing axis calculated using the correlation function (Equation 1).*

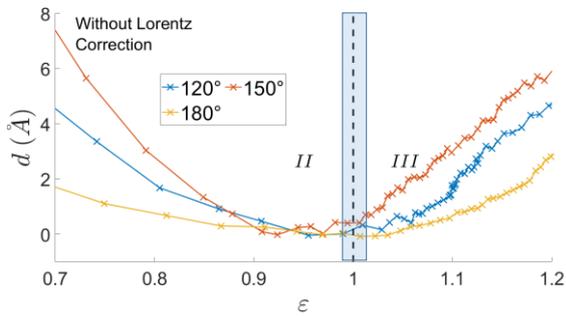

*Figure 10 Evolution of the correlation distance along the drawing axis calculated without the Lorentz correction at large strains (strong orientation).*

## 3.3 Microscopic phenomena evidenced by the WAXS analysis

### 3.3.1 α–β transition and chain orientation at large strains

All the reflections present in the patterns that were measured at the initial state ($\varepsilon = 0$) were successfully indexed using the unit cell parameters of the PBT α form given by Huo et al. [Huo (1992)] in a study where the temperature dependence of the parameters is taken into account (see Table 1 for the reflections visible on the pattern of Figure S5a, $T = 150°C$, distance sample/detector: 726 mm).

| Reflections | (001) | (011) | (010) | (111) | (110) | (100) |
|---|---|---|---|---|---|---|
| Lattice spacing measurement (Å) | 9.92 | 5.66 | 5.20 | 4.38 | Not measurable | 3.85 |
| Lattice spacing from Huo et al. at 150°C (Å) | 9.75 | 5.64 | 5.19 | 4.38 | 4.27 | 3.85 |

*Table 1 Lattice spacings at 150°C for the initial state ($\varepsilon = 0$, from Figure S5a): Measurements and calculated values using the PBT unit cell parameter given by Huo et al. [Huo (1992)] at 150°C (a=4.95 Å, b=6.09 Å, c=11.92 Å, α=100.9°, β=116.2° and γ=110.8°). The (110)reflection appears like an inconspicuous shoulder in the (111)peak and the corresponding lattice spacing is not measurable.*

In Figure S5b, we show an example of X-ray scattering pattern obtained now at large strain ($\varepsilon = 1.2$). The pattern reveals a strongly oriented morphology where the intensities associated with each reflection becomes concentrated at given azimuthal angles. We calculated the lattice spacings by measuring the peak positions in the pattern parts where the signal is concentrated. The results are gathered in Table 2. It was no longer possible to index satisfactorily these reflections with the unit cell parameters of the α triclinic structure (see table 1 line 3) but a very good agreement was found using those of the β triclinic structure that are given in Desborough et Hall [Desborough (1977)] (Table 2). These unit cell parameters were measured in the case of a highly oriented polymer in the β form. They were demonstrated to be more efficient to fit the experimental data than others that had been previously published [Yokouchi (1975)]. Unfortunately, the slight dependence of the unit cell parameters on temperature was not studied in the Desborough et al. article.

| | (001) | (011) | (010) | (111) | (110) | (100) |
|---|---|---|---|---|---|---|
| Lattice spacing measured (Å) | 10.37 | 5.59 | 5.18 | 4.04 | Not measurable | 3.76 |
| Lattice spacing from Desborough Hall (Å) | 10.31 | 5.59 | 5.17 | 4.09 | 3.91 | 3.73 |

*Table 2 Lattice spacings at $\varepsilon = 1.2$ : Measurements (150°C, Figure S5b) and calculated values using the PBT β unit cell parameters given by Desborough et Hall. [Desborough (1977)]. (a=4.73 Å, b=5.83 Å, c=12.90 Å, $\alpha$=101.9°, $\beta$=119.4° and $\gamma$=105.1°). The (11 0)) reflection is too weak to allow measurements. For each reflection, the peak positions were obtained at azimuthal angles where the intensities are maximum in the Figure S5b pattern.*

Our results suggest that the α–β transition indeed takes place during the deformation process. At the end of the test, the strong intensity concentrations in the WAXS patterns correspond to the β form that can then be considered as predominant at large strains.

By WAXS analysis of fiber patterns corresponding to the PBT β triclinic morphology, Yokouchi et al. found that the average direction (*c* axis of the triclinic unit cell) of the chains in crystals are along the drawing axis [Yokouchi (1975)]. In Figure S5b and S8 corresponding to large strains, it can be checked that the intensities corresponding to the (hk0) reflections—those of the planes that contain the chain direction—are concentrated along the vertical axis, axis perpendicular to the drawing direction ($\varphi$=90°). This confirms that the chain average directions in crystals are well along the drawing axis at large strains. This is not exactly true for other oriented polyesters like PET for which a 5° shift can exist between the chains and the drawing axis [Daubeny (1954)]. In the following, we will specifically analyze the evolution of the (001) reflection during the deformation process, in particular to follow the α–β transition.

### 3.3.2 Evolution of the (001) reflection during deformation

Examples of X-ray patterns showing the evolution of the (001) reflection as a function of strain are shown in Figure 11 (120°C), Figure S9 (150°C) and Figure S10 (180°C). At the initial state ($\varepsilon = 0$), the intensity is uniform along the diffraction circular ring. At larger strains, two features characterize the change of shape affecting this ring. Firstly, the intensity is no longer uniformly distributed along the ring, which is due to crystal reorientations that occur during the lamellar-fibrillar transition. Secondly, the ring is no longer perfectly circular. This means that the apparent (001) lattice spacing depends on the azimuthal angle $\varphi$, which will be interpreted by analyzing how the $\alpha-\beta$ transition depends on the crystal orientation.

### 3.3.3 Reorientation of the (001) reflection

In Figure 11, Figures S9 and S10, the reorientation of the (001) reflection is clearly visible at large strains ($\varepsilon = 1.1$). The intensity is then concentrated around a maximum situated at ($\varphi = 38°$). A good agreement was found with the value ($\varphi = 37°$) that can be calculated using the unit cell parameters of the $\beta$ triclinic form [Desborough (1977)], and assuming that the $c$ axis is along the drawing direction.

In order to evaluate the crystal reorientation rate, we have calculated the Herman factor: $H^{001} = \left(3\left\langle \cos^2\varphi \right\rangle - 1\right)\big/2$. When the lattice orientation is uniformly distributed: $H^{001} = 0$. As previously mentioned, when the chains in crystals are all oriented along the drawing direction, the final azimuthal angle value is approximately 38°, and the expected value for the Hermann factor should be $H^{001} \approx 0.42$. The evolution of $H^{001}$ is shown in Figure 12 for the three test temperatures. The three stages of the deformation process — separated by the same temperature-independent thresholds — can again be identified on the $H^{001}(\varepsilon)$ curves that are very similar to those corresponding to $A$. During strain range $I$ ($\varepsilon < 0.3$), the lamellar morphology is approximately preserved and $H^{001}$ slightly decreases. As previously mentioned, this is maybe due to the development of a "kink morphology" in the spherulite equatorial parts. During, the transition between the lamellar and fibrillar morphologies (strain range $II$, $0.3 < \varepsilon < 1$), $H^{001}$ strongly increases. This means that the chains in the crystals following the fragmentation/melting mechanism of lamellae are predominantly oriented along the drawing direction. During strain range $III$, crystal reorientation continues but the $H^{001}(\varepsilon)$ slope is slightly reduced. If the tensile test were to be continued, $H^{001}$ would likely increase to the 0.42 value that corresponds to fully oriented crystals.

### 3.3.4 Change of the (001) plane spacing along the ring in link with the α–β transition

To visualize clearly the lattice spacing changes along the (001) ring, we have plotted an identical quarter circle in dotted lines on all the patterns corresponding to the same temperature (Figure 11, S9 and S10). At $\varepsilon = 0$, this quarter circle corresponds to the position of the intensity maxima along the ring. At $\varepsilon = 0.3$, the intensity maxima are respectively situated inside and outside the reference quarter circle in the horizontal ($\varphi \approx 0$) and vertical ($\varphi \approx 90°$) pattern parts. This means that the (001) plane spacing is increased when the normal to these planes is close to the tensile axis while it is decreased when it is perpendicular to this direction. At $\varepsilon = 0.7$, the intensity maximum of the (001) reflection along the horizontal axis has moved even closer to the beam stop (ring center). In Figure 13, we have plotted $d_{001}(\varphi)$ for different strains at 120°C, as well as the expected $d_{001}$ values for the α and β forms. The $d_{001}(\varphi)$ curves were obtained by finding the $q$ position of the peak associated with the (001) reflection at a given value of the azimuthal angle $\varphi$. Because the corresponding $q$ values are close, and because as it will be shown thereafter these two phases are intrinsically slightly deformable along the normal to the (001) planes, it was not possible to separate the two peaks that are associated with the α and β forms. Therefore, the continuous variations of $d_{001}$ between the expected values for these two phases that are observed roughly for $\varphi < 50°$—when the macro-elongation along the normal to the lattices is positive (see appendix A)—show that these two crystalline forms can coexist for crystals with the same orientation. The β forms becomes predominant when $\varphi \rightarrow 0$, that is to say when the stress and macro-elongation along the normal to the 001 lattices becomes maximum. For $\varphi = 0°$, $d_{001}$ is even larger than the value (10.31 $Å$) corresponding to the β phase. The unit cell parameters that we used for this triclinic phase were determined on a strongly oriented structure [Desborough (1977)], which corresponds to $\varphi = 38°$ for the (001) reflection, an orientation for which the stress ($\sigma \cos^2 \varphi$) along the normal to the considered lattices is significantly reduced compared to $\varphi = 0$. After the formation of the β structure, $d_{001}$ remains therefore slightly extensible when the strain (and stress) is applied along the normal to the (001) lattices with a maximum possible value around 10.55 $Å$. For lattice orientations subjected to negative macro-elongation along their normal (roughly $\varphi > 65°$, see appendix A), $d_{001}$ becomes a bit lower than the value corresponding to the α triclinic structure (9.92 $\mathring{A}$) with a minimum value of about 9.85 $\mathring{A}$. There again, the α crystals are slightly deformable even in the absence of any possible phase change.

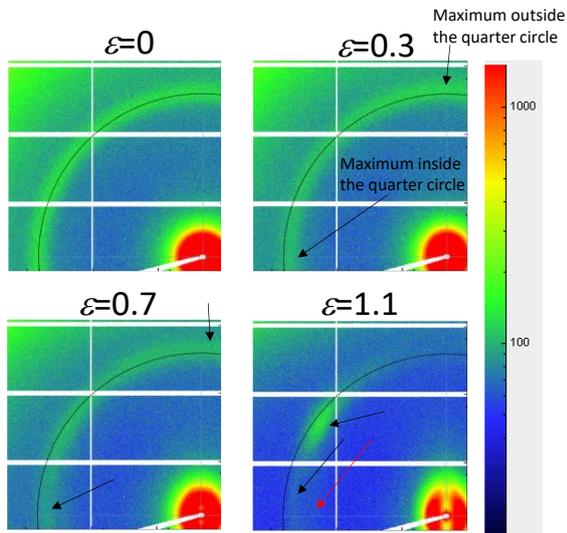

*Figure 11 Examples of patterns showing the evolution of the (001) reflection shape (T=120°C). Red arrow: unexpected intensity reinforcement.*

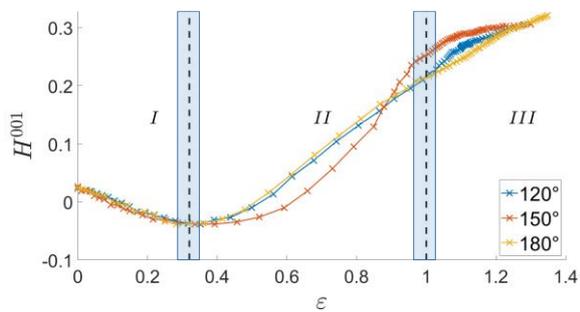

*Figure 12 Evolution of the Herman orientation factor.*

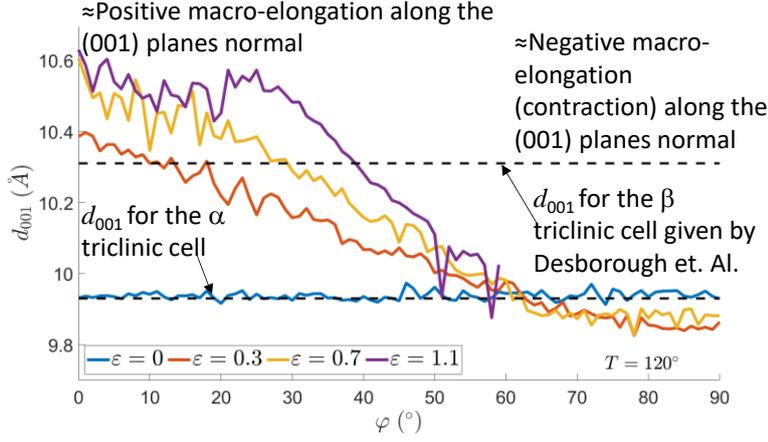

*Figure 13* $d_{001}$ *versus the azimuthal angle along the (001) reflection ring at different strains. At* $\varepsilon = 1.1$ *, the ring intensity becomes very weak for* $\varphi > 60°$ *and* $d_{001}$ *cannot be measured.*

### 3.3.5 Analysis of the α–β transition as a function of the orientation

In the following, we use the formalism presented in appendix A to analyze further and quantitatively the relationship between the macro-scale deformation and the evolution of $d_{001}$ in link with the α→β transition. During strain range *I*, i.e. before the yield strain, the integrity of the lamellar morphology is considered as preserved and no breakage in crystals occurs. We therefore assume that the changes of the (001) lattice normal directions with respect to the drawing axis are the same as that of the macroscopic directions that coincide at the initial state. This means that $\psi(\psi_0, \varepsilon) = \psi_L(\psi_0, \varepsilon)$ with the notations defined in sections 2.2.2 and appendix A. It should also be noted that the orientation of the (001) lattices only slightly change during the α→β transition, from 35° to 36.5° with respect to the chain direction (*c*).

Using the formalism presented in the appendix A, we can calculate $\psi(\psi_0, \varepsilon)$ (see Figure A1 of appendix A), and using $\varphi \approx \psi$ (section 2.2.2) extract the local elongation in crystals from patterns: $e_{crys}(\psi_0, \varepsilon) = \left[d_{001}^{\psi}(\varepsilon) - d_{001}^{\psi_0}(0)\right] \Big/ d_{001}^{\psi_0}(0)$. $d_{001}^{\psi_0}(0)$ is the (001) lattice spacing in a crystal oriented so that the normal to the (001) plane makes an initial angle $\psi_0$ with respect to the drawing axis. $d_{001}^{\psi}(\varepsilon)$ is the (001) lattice spacing during deformation for the same crystal when the angle between the normal and the drawing axis has changed from $\psi_0$ to $\psi$. The objective is to analyze the link of $e_{crys}(\psi_0, \varepsilon)$ with the macro-elongation $e_{macro}(\psi_0, \varepsilon) = \left[\ell^{\psi}(\varepsilon) - \ell^{\psi_0}(0)\right] \Big/ \ell^{\psi_0}(0)$. As detailed in appendix A, $\ell^{\psi_0}(0)$ and $\ell^{\psi}(\varepsilon)$ are the initial and current lengths of an element defined on the macro-scale that makes an initial angle $\psi_0$, and a current angle $\psi$, with respect to the drawing axis, as the normal to the considered (001) lattices. The evolution of $e_{crys}$ versus $\varepsilon$ is plotted for various $\psi_0$ initial angles in Figure 14 (120°C), S11 (150°C) and S12 (180°C). When the macro-elongation $e_{macro}$ is positive during strain range $I$ ($\varepsilon < 0.3$), plain circles symbol are used ($\psi_0$=0°, 20°, 40° and 50°, Figure A2 in appendix A) whilst it is plain triangles otherwise ($e_{macro} < 0$; $\psi_0 = 80°$ and 90°, Figure A2). For $\psi_0 = 65°$, $e_{macro} \approx 0$ during stage I (Figure A2) and we used plain squares for the plot.

The comments about the curves representing $e_{macro}$ against $\varepsilon$ (Figures 14, S11 and S12) are consistent with those corresponding to Figure 13. When $e_{macro}$ is positive, $e_{crys}$ is also positive (plain circle curves), and the larger $e_{macro}$ ($\Leftrightarrow$ the smaller $\psi_0$) the larger $e_{crys}$. On the other hand, if $e_{macro}$ is negative, $e_{crys}$ is also negative (plain triangle curves). When $e_{macro}$ is very close to 0, $e_{crys}$ is also very close to 0 (plain square curve). For further analysis, we plotted $|e_{crys}|$ against $|e_{macro}|$ (Figure 15: 120°C, Figure S13: 150°C and Figure S14: 180°C). Strikingly, if $\psi_0 \leq 50°$, all the curves fall on the same master curve. This means that if $e_{macro} > 0$, the average elongation ($e_{crys}$) in crystals along the (001) lattice normal only depends on the macro elongation ($e_{macro}$) along this direction. At 120°C, $e_{crys}$ approximately reaches the value corresponding to the β structure when $e_{macro} \approx 0.3$., which corresponds to $d_{001} = 10.31 \mathring{A}$ or $e_{crys} \approx 0.04$. Overall, the relative proportion of each of the two phases appears to be controlled by the macro-elongation applied along the normal to the (001) lattices. For $e_{macro} = 0.3$, $e_{macro}$ is respectively 7.2, 8.0 and 10.3 times larger than $e_{crys}$ at 120°C (Figure 15), 150°C (Figure S13) and 180°C (Figure S14). The deformation mainly takes place in the interlamellar amorphous regions. At 150°C and 180°C, $e_{crys}$ does not reach the value corresponding to the β form at $e_{macro} = 0.3$. However, the significant increase of $e_{crys}$, average value between the α and β forms (because the corresponding peaks are not separated), shows that the phase transition has well begun. If $e_{macro} < 0$, the dependence of $e_{crys}$ on $e_{macro}$ differs from that observed when $e_{macro} > 0$, the $e_{crys} / e_{macro}$ ratio is smaller. The $e_{crys}$ decrease is then only due to the compressibility of the α crystals, no phase transition is involved.

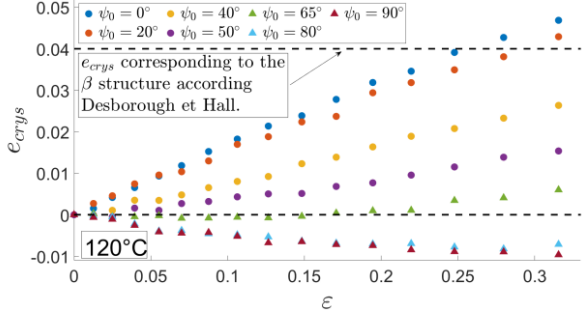

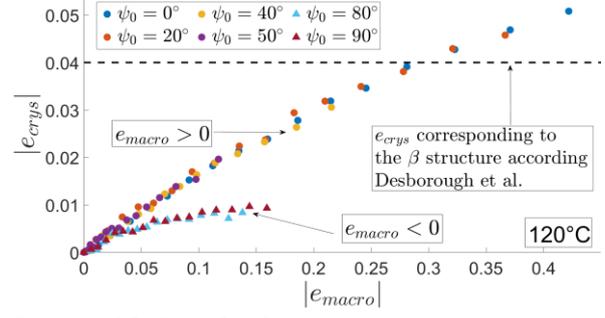

*Figure 14 Evolution of the local elongations versus $\varepsilon$. $\psi_0$ is the initial angle between the normal to the (001) lattices and the drawing axis (T=120°C).*

*Figure 15 Local elongations versus macro-elongations. Results provided in absolute values. (T=120°C).*

## 3.4 Evidences of the development of a mesomorphic phase at large strains

At larger strains and for small $\varphi$ angles, we detected a new weak and unexpected peak at the position indicated by a red arrow in Figure 11 or Figure S5b. In Figure S15, we show the pattern corresponding to the image of Figure 11 at $\varepsilon = 1.1$ but with color scale modified to make the new peak easier to observe. The appearance and growth of this new peak can also be analyzed by plotting the $I(q)$ intensity profiles measured at different strains for $\varphi = 0°$ (Figure 16). On these profiles, a deconvolution procedure was implemented to separate the new peak from that associated with the (001) reflection. The new peak (denoted SP) can clearly be extracted from the noisy intensity signal for strains larger than $\varepsilon = 0.3$ or $\varepsilon = 0.4$. Next, it grows as the strain increases and its maximum becomes higher than that of the (001) reflection at the end of the test ($\varepsilon = 1.28$). We plotted in the same figure (Figure S16) the correlation distance evolutions for the two peaks: $d_{SP}$ (New Peak) and $d_{001}$. $d_{SP}$ starts to be measurable approximately at $\varepsilon \approx 0.4$, and then remains approximately constant, even if a slight increase trend can be detected at the test ends, precisely at the strain threshold $\varepsilon = 1$ when the fibrillar morphology is established and the polymer chains taut. We found the following values for $d_{SP}$ at 120°C, 150°C and 180°C: $11.9\ \mathring{A}$, $11.9\ \mathring{A}$ and $12.0\ \mathring{A}$ respectively. At each temperature these values were averaged in the $\varepsilon = \begin{bmatrix} 0.6 & 0.8 \end{bmatrix}$ interval where the new peak is accurately extracted from the noise.

The smectic peaks that have already been observed in amorphous polyester subjected to stretching below $T_G$ are situated along the drawing axis [Song(2000),Konishi (2010), Kawakami(2005), Sago(2014), Jakeways (1996)]. In the case of PBT, this peak corresponds to a $11.69\,\mathring{A}$ correlation distance [Song(2000), Konishi (2010), Kawakami(2005), Sago(2014)]. The peak appearing at large strains and $\varphi \approx 0$ (drawing axis) in Figure 16 can therefore be identified as the smectic phase already reported for PBT. The slight discrepancy in correlation distance —$\approx 11.9 \mathring{A} - 12\,\mathring{A}$ in the 120°-180° range instead of $11.69\,\mathring{A}$ at room temperature — can be attributed to thermal expansion. Of note, this phase is there observed during stretching of an initially crystallized polymer, and not starting from an amorphous polymer. Moreover, this phase is observed there well above $T_G$ and in a range of temperature, approaching the melting point.

In order to analyze the development characteristics of the smectic phase, we have plotted the area of the SP peak divided by specimen thickness (*h*) versus $\varepsilon$. (Figure 17: 120°C, Figure S17: 150°C and Figure S18 180°C). The normalization by *h* allows for accounting the intensity signal decrease due to thickness reduction.

In strain range *I*, the peak area is close to zero and below the detection limit. At strain threshold $\varepsilon \approx 0.3$, the SP area begins to increase This suggests that the polymer orientation, which strongly increases during strain range II (see Figure 7 and 12), could be the main factor leading to the appearance of the smectic phase. However, it is difficult to assess quantitively the development of the smectic phase from Figure 17 because reflections oriented along the draw axis may no longer satisfy the Bragg condition, which would lead to decreasing the peak intensity at φ = 0, meaning that quantification of the smectic development may be underestimated. At the second strain threshold ($\varepsilon \approx 1.0$), the slope slightly increases.

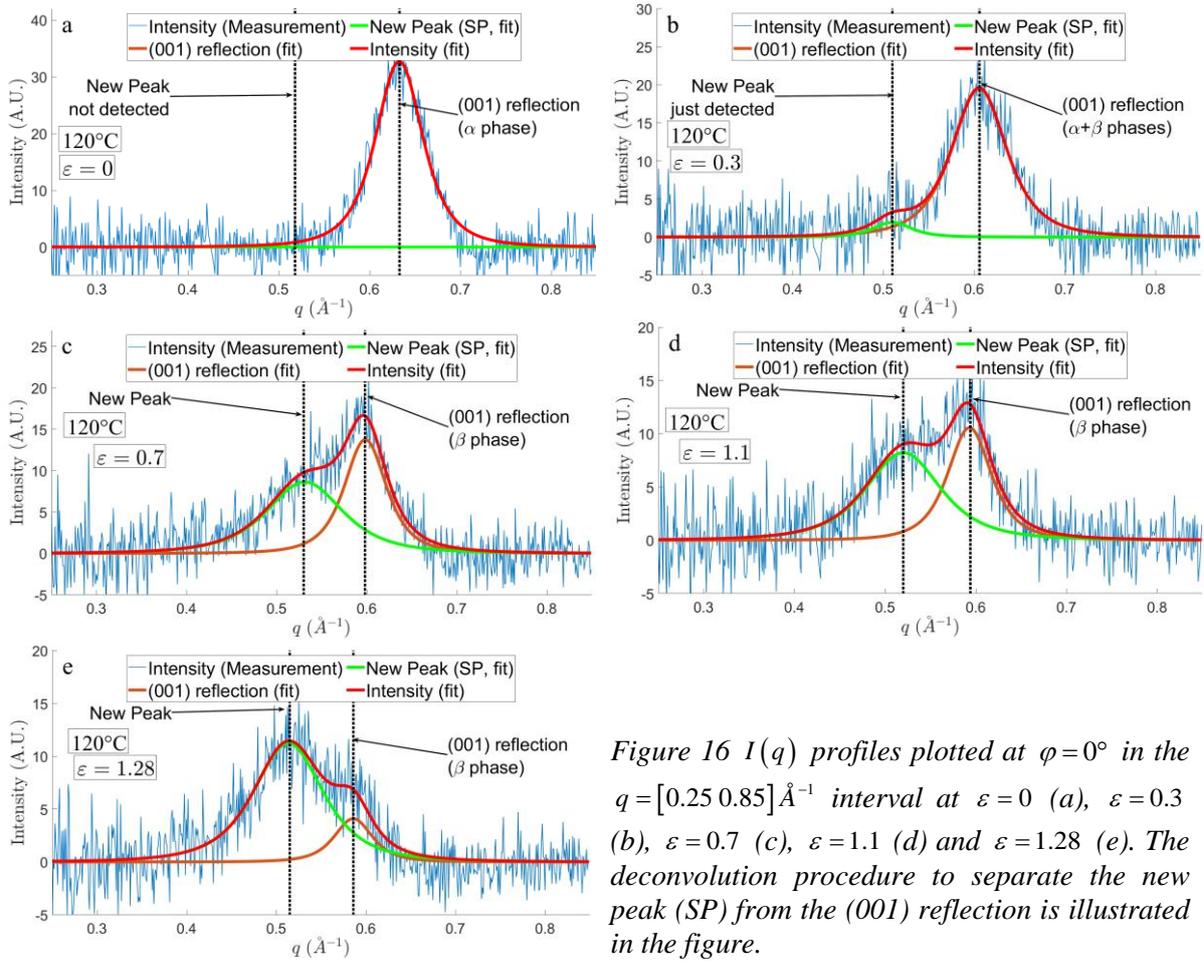

*Figure 16 $I(q)$ profiles plotted at $\varphi = 0°$ in the $q = [0.25\,0.85]\,Å^{-1}$ interval at $\varepsilon = 0$ (a), $\varepsilon = 0.3$ (b), $\varepsilon = 0.7$ (c), $\varepsilon = 1.1$ (d) and $\varepsilon = 1.28$ (e). The deconvolution procedure to separate the new peak (SP) from the (001) reflection is illustrated in the figure.*

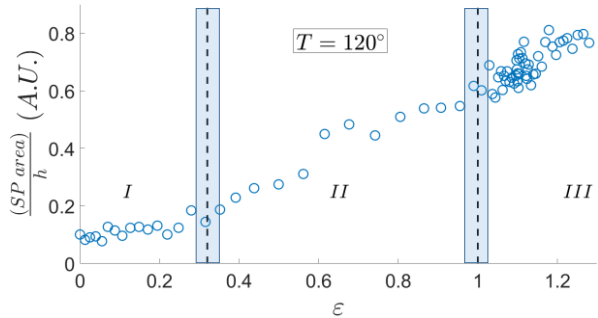

*Figure 17 Area of SP (new peak associated with the smectic phase) normalized by the specimen thickness (h) versus $\varepsilon$ (120°C).*

## Conclusion

Synchrotron X-ray scattering experiments were performed to study the deformation mechanisms in

PBT. Thanks to the design of a heating device optimized for the beamline, a combination of SAXS-

WAXS measurements were *in-situ* obtained during tensile tests carried out at 120°C, 150°C and 180°C

on a crystallized and initially non-oriented PBT. Specific procedures were implemented to extract from the scattering patterns physical quantities associated with the material's deformation mechanisms. The deformation process was found to occur in three stages, separated by temperature-independent macroscopic strain thresholds. Another original result concerns the α→β transition studied from the evolution of the (001) lattice spacing and shown as fully governed by the macroscopic elongation along the normal to these lattices. Finally, the development of a smectic phase was observed at large strains above $T_G$, in conditions never reported before.

## Appendix A: Analysis of the macro strain state during tensile testing

We use classical continuum mechanics results to determine the strain state at the macro-scale in the specimen center.

Let $\ell^{\psi_0}$ be a small length element along a direction corresponding to the $\overrightarrow{n_{\psi_0}}$ unit vector that makes an angle $\psi_0$ with the drawing axis at the initial state ($\varepsilon = 0$). $\ell^{\psi_0}$ is considered to be associated with the macro-scale, which means that the microstructure is assumed as averaged. During deformation ($\varepsilon \neq 0$), $\ell^{\psi_0}$ becomes $\ell^{\psi}$ and is currently along the $\overrightarrow{n_{\psi}}$ unit vector that makes an angle $\psi(\psi_0, \varepsilon)$ with the drawing axis. The macro-elongation is defined by:

$$e_{macro}\left(\psi_0, \varepsilon\right) = \frac{\ell^{\psi}\left(\varepsilon\right) - \ell^{\psi_0}\left(0\right)}{\ell^{\psi_0}\left(0\right)}$$

This appendix briefly recalls how to calculate $\psi$ and $e_{macro}$ as a function of $\psi_0$ and $\varepsilon$. Assuming a uniaxial stress state, transverse anisotropy around the tensile axis and an isochoric deformation process (Poisson's ratio $\nu = 0.5$), the Hencky strain tensor in the specimen central point, where the X-ray beam passes through the specimen thickness, is given by:

$$\underline{\underline{\varepsilon}} = \begin{pmatrix} \varepsilon & 0 & 0 \\ 0 & -\dfrac{1}{2}\varepsilon & 0 \\ 0 & 0 & -\dfrac{1}{2}\varepsilon \end{pmatrix}.$$

Axis 1 corresponds to the drawing axis. The specimen thickness is $h = h_0 \exp\left(\dfrac{-\varepsilon}{2}\right)$. $h_0$ ($4\,mm$) is the initial thickness. The transformation gradient tensor writes:

$$\underline{\underline{F}} = \begin{pmatrix} \exp(\varepsilon) & 0 & 0 \\[2mm] 0 & \exp\left(-\dfrac{\varepsilon}{2}\right) & 0 \\[2mm] 0 & 0 & \exp\left(-\dfrac{\varepsilon}{2}\right) \end{pmatrix}$$

Using classical results of finite strain theory, it can be shown that that:

$$\overrightarrow{n_\psi}\left(\psi_0, \varepsilon\right) = \frac{\underline{\underline{F}}\,\overrightarrow{n_{\psi_0}}}{\sqrt{\overrightarrow{n_{\psi_0}}.\underline{\underline{F}}\,\underline{\underline{F}}^T\,\overrightarrow{n_{\psi_0}}}} \qquad\qquad \text{Equation 3}$$

From the $\overrightarrow{n_\psi}\left(\psi_0, \varepsilon\right)$ components, it is easy to obtain $\psi\left(\psi_0, \varepsilon\right)$. The macro-elongation is given by:

$$e_{macro}\left(\psi_0, \varepsilon\right) = \sqrt{\overrightarrow{n_{\psi_0}}.\underline{\underline{F}}\,\underline{\underline{F}}^T\,\overrightarrow{n_{\psi_0}}} - 1 \qquad\qquad \text{Equation 4}$$

In Figure A1, we show the evolution of $\psi$ during a tensile test. Logically, for $\psi_0 = 0°$ or $\psi_0 = 90°$, $\psi$ remains constant throughout the test. In the $0° < \psi_0 < 90°$ interval, $\overrightarrow{n_\psi}$ becomes close to the tensile axis and $\psi \to 0$ when $\varepsilon$ increases. In Figure A2, we show the evolution of the macro-elongation ($e_{macro}$) corresponding to length elements of initial orientation $\psi_0$ with respect to the tensile axis. It can be checked that if $\psi_0 < 54,7°$, the element lengths are stretched throughout the test ($e_{macro} > 0$), and $e_{macro}$ always increases during the test. As expected, the smaller $\psi_0$, the larger $e_{macro}$. In the $54.7° < \psi_0 < 90°$ range, the length element undergoes an initial contraction phase ($e_{\psi_0} < 0$). It can be noted that for $\psi_0 = 65°$, $e_{macro}$ remains very close to zero until $\varepsilon \approx 0.6$. In the $\varepsilon$ strain range corresponding to this study ($0 \le \varepsilon \le 1.3$, see Figure 2), $e_{macro} < 0$ for $\psi_0 = 80°$ and off course for $\psi_0 = 90°$.

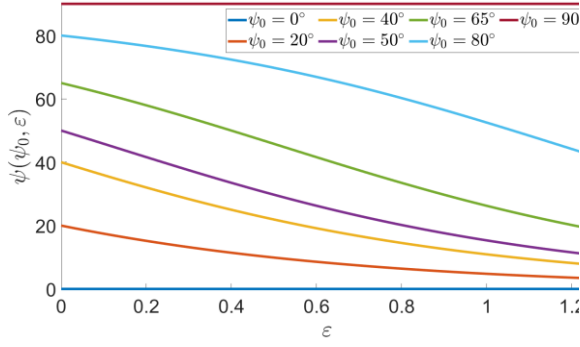

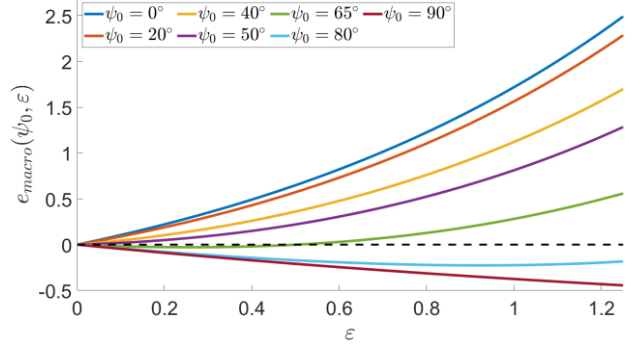

*Figure A1 Deformation induced evolution of the angle $\psi$ between the drawing axis and a direction initially making an angle $\psi_0$ with the drawing axis*

*Figure A2 Evolution of the macro-elongation $e_{macro}(\psi_0, \varepsilon) = \dfrac{\ell^{\psi}(\varepsilon) - \ell^{\psi_0}(0)}{\ell^{\psi_0}(0)}$ for a length element $\ell^{\psi_0}$ initially making an angle $\psi_0$ with respect to the drawing axis*

## Acknowledgments

The authors would like to thank the ICEEL Carnot Institute, the EMPP research department of Lorraine University, the Grand Est Region and ADEME (French agency for ecological transition) for their financial support.

We are also very grateful to Diego Ciardi for helping prepare and carry out these experiments.

Preliminary experiments have been performed at the DND-CAT 5ID-D beamline at Argonne National Laboratory under Contract DE-AC02-06CH11357. We thank Steven J. Weigand for performing these measurements.

# Appendix B Supplementary Data

## SD1 Validity of the isochoric deformation process in presence of the α–β transition?

We calculated that the change of the unit cell volume associated with the α–β transition is 5%. As in the rest of the paper, we used unit cell parameters given by Huo et al. [Huo(1992) for the α form and Desborough et al. [Desborough(1977)] for the β form. We had to take the α unit cell parameters at room temperature because we did not find any published study giving the dependence of the β unit cell parameters on temperature. Obviously, the comparison of the unit cell volume must be performed at the same temperature to avoid the effect of thermal expansion. In addition to those corresponding to the β form, Desborough et al. also gives the unit cell parameters corresponding to the α form at room temperature. Using this sole paper for the two phases, we found that the change in unit cell volume during the α–β transition is slightly smaller: 4.4%.

The crystallinity is in the 30-35% range and the volume strain specifically associated with a complete transformation from α form to β form is therefore approximately 1.7% (33%×5%). At the first critic strain identified in this work ($\varepsilon \approx 0.3$), we can estimate that the intensity along the (001) ring remains approximately uniform (see Figure 12) and that about a half of the α crystals have undergone the polymorphic transformation (see Figure 13). Using the equation giving the volume strain: $\varepsilon_v = \varepsilon + 2\varepsilon_t$ ($\varepsilon_t$: transverse strain), we can roughly estimate the errors due to the application of the isochoric deformation assumption. The error is 2.8% on $\varepsilon_t$ (-0.146 instead of -0.15) and 0.86% on the true stress ($\sigma_V = \sigma_N \exp(2 \times 0.146) = 1.3391 \times \sigma_N$ instead of $\sigma_V = \sigma_N \exp(0.3) = 1.3499 \times \sigma_N$). At the second critic strain ($\varepsilon \approx 1.0$), the α–β transition is finished and the errors on $\varepsilon_t$ and $\sigma_V$ are both approximately of 1.7%.

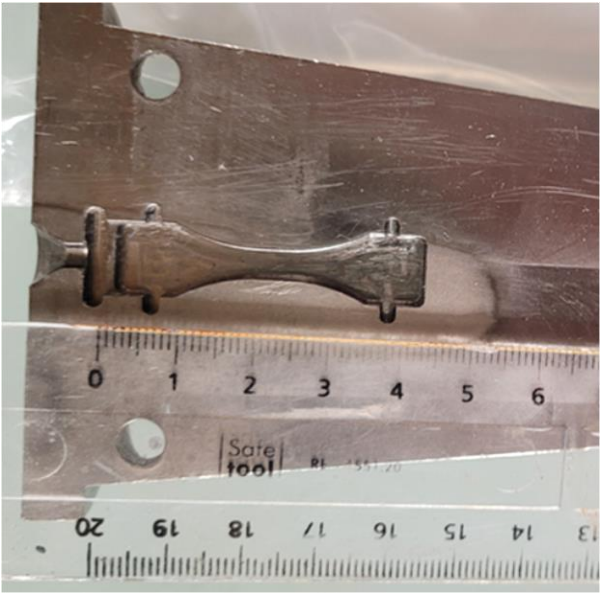

Figure S1 Photography of the specimen mold.

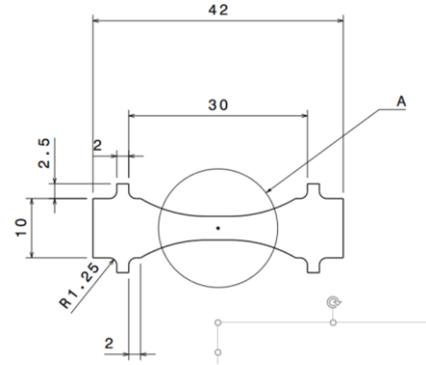

Figure S2 Specimen dimensions.

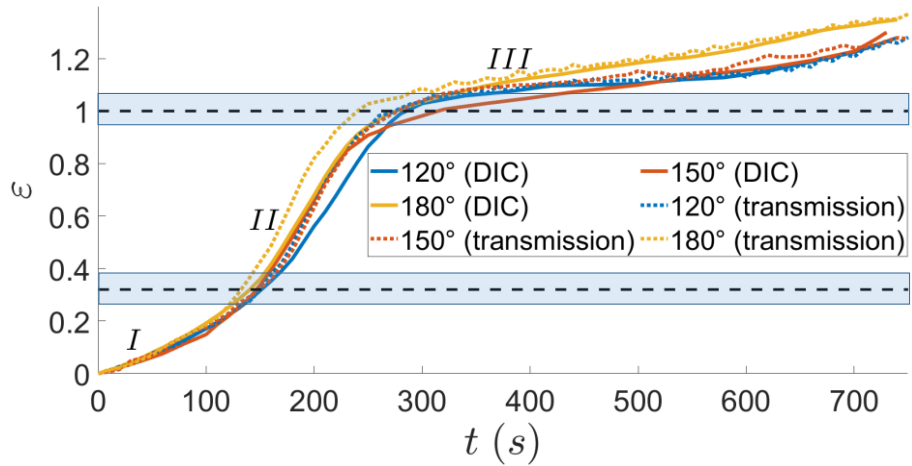

Figure S3 Evolution of true strain ( $\varepsilon(t)$ ) measured in the specimen central cross-section. Comparison between the DIC measurements and those obtained independently during the in situ X Ray scattering experiments from the transmission signal. Using the beer Lambert-Law and the assumption of isochoric deformation process, it is easy to check that: $\varepsilon(t) = -2\ln\left(\dfrac{\ln T(t)}{\ln T(0)}\right)$. $T$ (transmission) is given by $T = I/I_0$ , measured intensity divided by intensity in absence of sample.

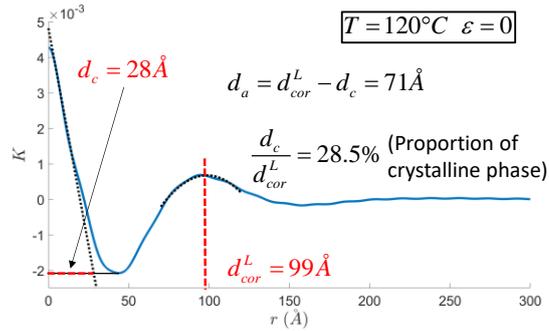

*Figure S4 Illustration of the procedure to obtain $d_c$ (thickness of lamellae), $d_a$ (thickness of amorphous layers) and the proportion of crystalline phase through the characteristics of $K(r)$ ([Strobl (1980)] for our PBT at 120°C and $\varepsilon = 0$.*

## SD2 Comments about the calculation and use of the correlation function $K(r)$

The measured $I_{meas}(q)$ intensity was fitted in the $q = [0.2\,0.35]\,Å^{-1}$ range by equation $I_{meas}(q) = \dfrac{A}{q^3} + B$.

At large $q$ values, the $I(q)$ intensity used in Equation 1 was obtained by replacing $I_{meas}$ by the fitted model ($\dfrac{A}{q^3} + B$). $B$ was next subtracted from the signal. This was aimed at reducing the measurement noise ($\approx B$) and the parasitic oscillations in $K(r)$.

The linear model used to obtain $d_c$ is calculated in the $r = [3\,8]\,Å$ interval. The determination of $d_{cor}^L$ was refined by fitting the data with a 2nd order polymeric function around the corresponding maximum. In that case $d_c$ corresponds to the thickness of lamellae because the proportion of crystal phase is well below 50% (DSC measured crystallinity: 34%). Overall, $d^{Lz}$ determined using the $K(r)$ correlation function was found to be slightly smaller than those obtained directly by finding the $q^2 I(q)$ maximum (see sections 2.2.2 and 3.3.2). The difference is always nearly the same. If we had used $d^{Lz}$ calculated through the correlation function in Figure 8, the curves would just have been slightly shifted.

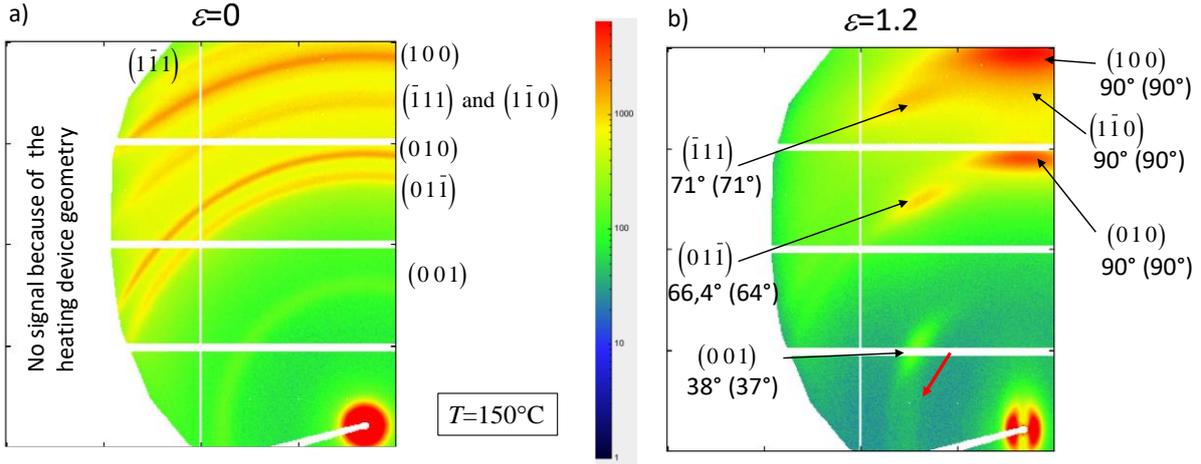

*Figure S5 Patterns measured for the 726 mm sample/detector distance at 150°C, a)  $\varepsilon = 0$  and b)  $\varepsilon = 1.2$. In Figure S5b, we indicated the azimuthal angle ( $\varphi$ ) corresponding to the position of each reflection maximum. Between brackets, the angle calculated for each reflection maximum using the unit cell parameters of the $\beta$ form and assuming that the chains are along the drawing axis. Red arrow: unexpected intensity reinforcement associated with the appearance of a new phase at large strains.*

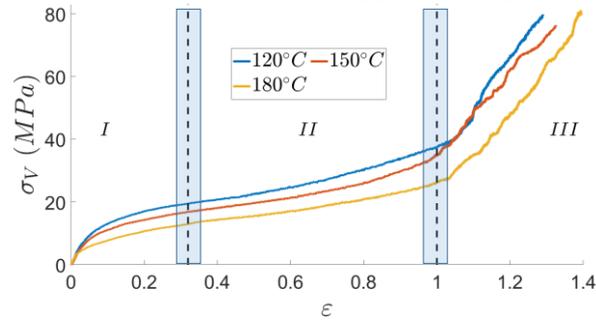

*Figure S6 True strain-true stress curves ( $\sigma_v(\varepsilon)$ ).*

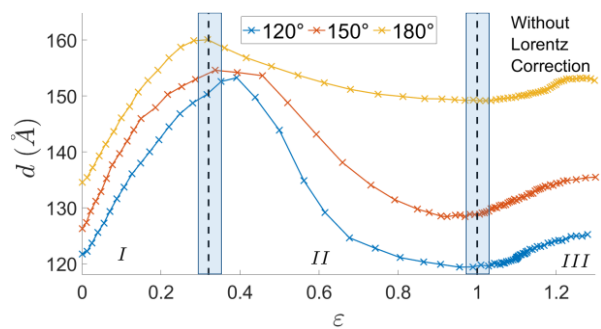

*Figure S7 Evolution of the correlation distance along the drawing axis calculated without using the Lorentz correction.*

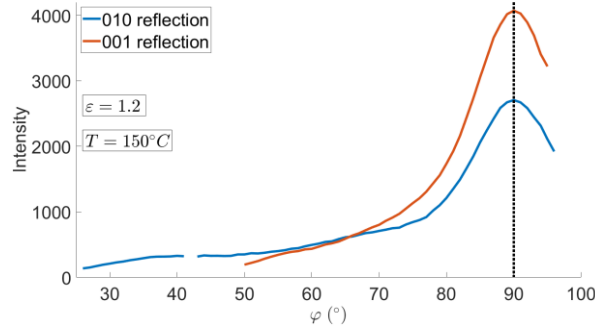

*Figure S8 Intensity profiles along the (010) and (001) reflections at $\varepsilon = 1.2$ from Figure S5b.*

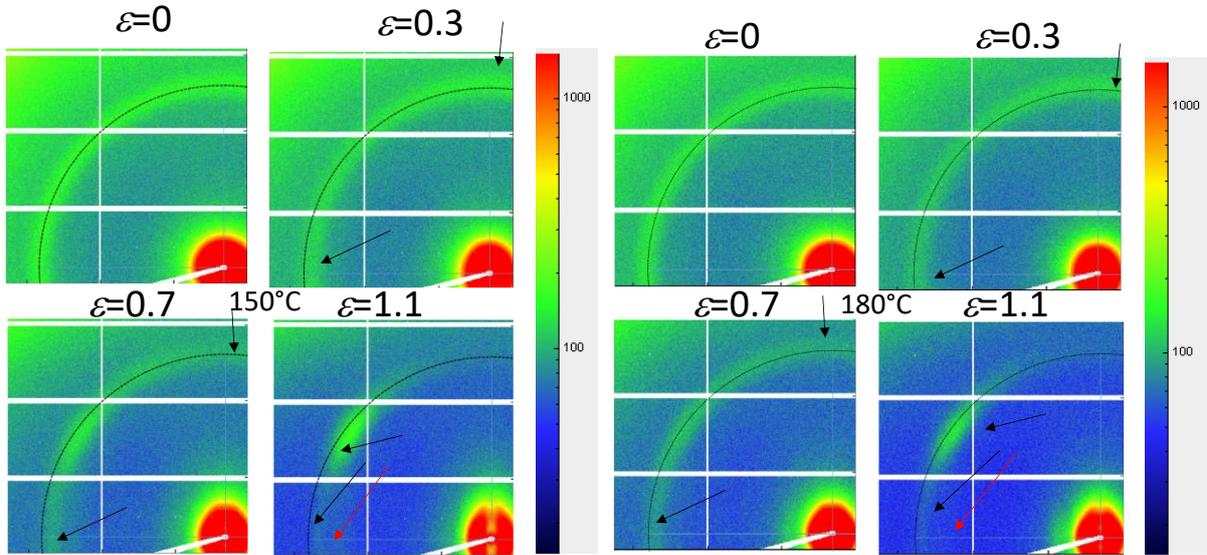

*Figure S9 Examples of patterns showing the evolution of the (001) reflection shape (T=150°C).*

*Figure S10 Examples of patterns showing the evolution of the (001) reflection shape (T=180°C).*

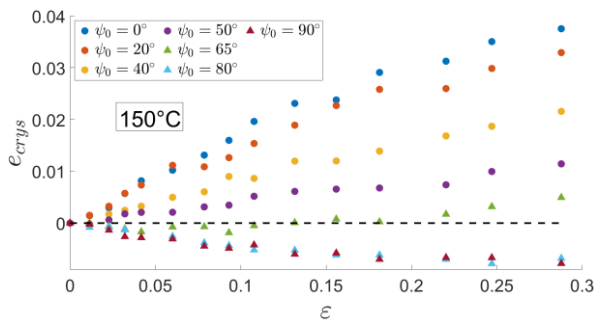

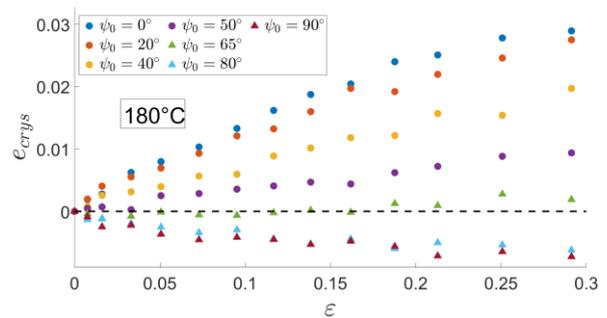

*Figure S11 Evolution of the local elongations versus $\varepsilon$. $\psi_0$ is the initial angle between the normal to the (001) lattices and the drawing axis (T=150°C).*

*Figure S12 Evolution of the local elongations versus $\varepsilon$. $\psi_0$ is the initial angle between the normal to the (001) lattices and the drawing axis (T=180°C).*

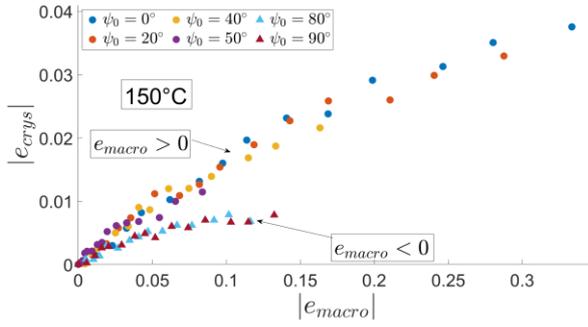

*Figure S13 Local elongations versus macro-elongations. Results provided in absolute values (T=150°C).*

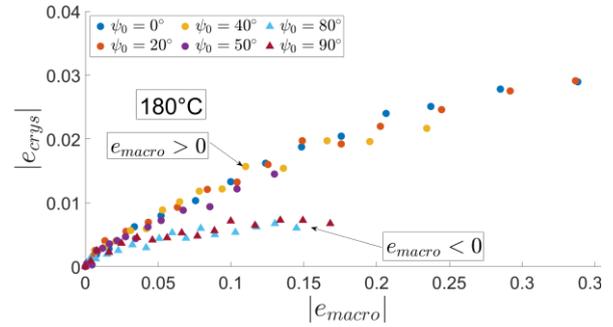

*Figure S14 Local elongations versus macro-elongations. Results provided in absolute values (T=180°C).*

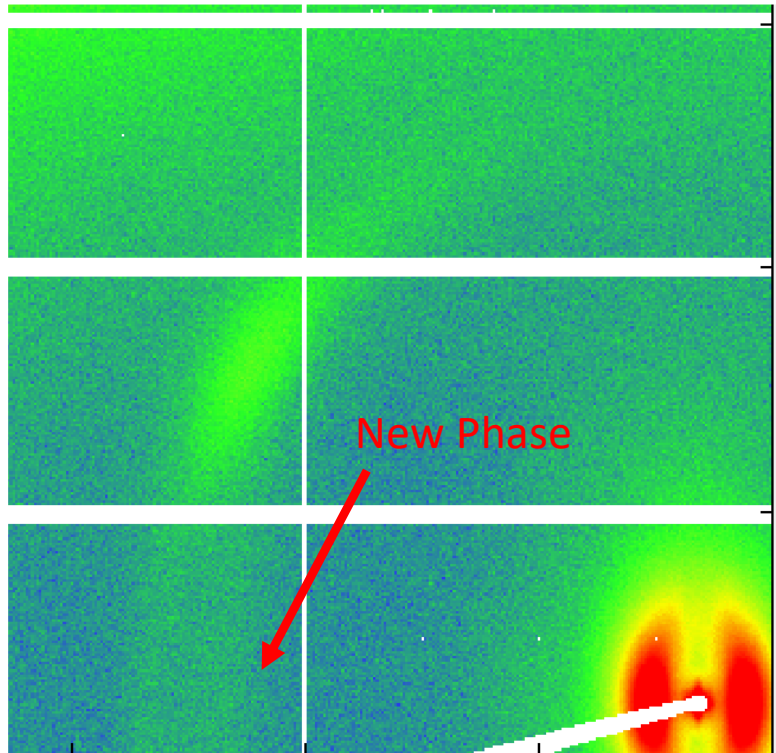

*Figure S15 Pattern measured at 120°C and $\varepsilon = 1.1$. Compared to Figure 13, the noise threshold was lowered to make the new peak associated with the new phase easier to observe.*

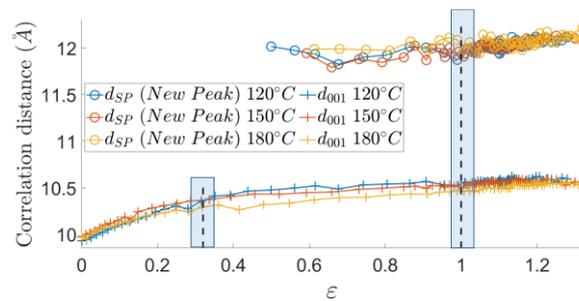

*Figure S16 Evolution of the correlation distance for the (001) reflection and the new peak corresponding to the smectic phase (obtained from the $I(q)$ profiles averaged in the $\varphi = \begin{bmatrix} 0 & 5 \end{bmatrix}°$ interval).*

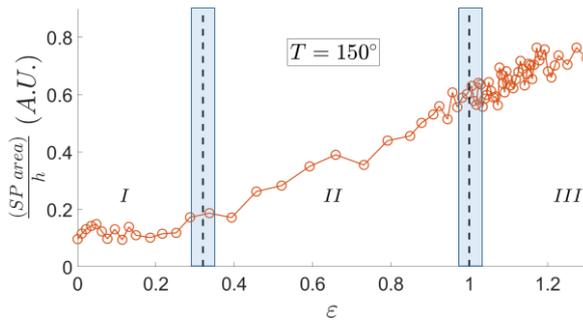
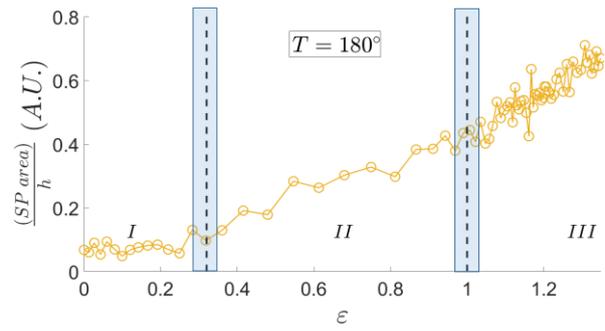

*Figure S17 Area of SP (new peak associated with the smectic phase) normalized by the specimen thickness (h) versus ε (150°C).*

*Figure S18 Area of SP (new peak associated with the smectic phase) normalized by the specimen thickness (h) versus ε (180°C).*